\documentstyle[11pt]{article}

\newcommand{\pslash}{P \hspace{-0.24cm} / \,}
\newcommand{\kslash}{k \hspace{-0.21cm} / \,}
\newcommand{\nslash}{n_{-} \hspace{-0.44cm} / \;\;}
\newcommand{\sslash}{S \hspace{-0.22cm} / \,}

\begin{document}

\title{\bf Parameterization of the quark-quark correlator of a 
spin-${\boldmath \frac{1}{2}}$ hadron}

\author{K.~Goeke, A.~Metz, and M.~Schlegel
 \\[0.3cm]
{\it Institut f\"ur Theoretische Physik II,} \\
{\it Ruhr-Universit\"at Bochum, D-44780 Bochum, Germany}}

\date{\today}
\maketitle

\begin{abstract}
\noindent
The general parameterization of the quark-quark correlation function for
a spin-$\frac{1}{2}$ hadron is considered.
The presence of the Wilson line ensuring color gauge invariance of the 
correlator induces structures that were not given explicitly in the 
existing literature.
In particular, the general form of the transverse momentum dependent
correlator entering various hard scattering processes is derived.
In this case two new time-reversal odd parton distributions appear at
the twist-3 level.
\end{abstract}

\noindent
{\bf 1.}~The purpose of this note is to provide the general structure of the
quark-quark correlation function of a spin-$\frac{1}{2}$ hadron,
\begin{equation} \label{e:corr1}
\Phi_{ij}(P,k,S \, | n_{-})  =
 \int \frac{d^4 \xi}{(2 \pi)^4} \, e^{i k \cdot \xi} \,
 \langle P,S \, | \, \bar{\psi}_j(0) \,
 {\cal W}(0,\xi| n_{-}) \, \psi_i(\xi) \, | \, P,S \rangle\, .
\end{equation}
The target state is characterized by its four-momentum
$P$ and the covariant spin vector 
$S$ $(P^2 = M^2, \; S^2 = -1 , \; P \cdot S = 0)$, while $k$ denotes the
momentum of the quark.
The Wilson line ${\cal W}(0,\xi| n_{-})$ guarantees color gauge invariance 
of the correlator, where the specific path of the gauge link will be given 
below. 
Several articles in the 
literature~\cite{ralston_79,mulders_95,goeke_03,bacchetta_04} are already
dealing with the general parameterization of $\Phi$, but none of them contains 
explicitly the complete decomposition.

The knowledge of the correlator in Eq.~(\ref{e:corr1}) is particularly 
useful in order to obtain the general form of the transverse momentum
dependent ($k_{T}$-dependent) correlator $\Phi(x,\vec{k}_{T},S)$, 
which enters the description of hard scattering processes like transverse
momentum dependent semi-inclusive DIS and the unintegrated Drell-Yan
reaction.
The connection between both objects is given by the relation
\begin{equation} \label{e:rel}
\Phi(x,\vec{k}_{T},S) = \int dk^{-} \, \Phi(P,k,S \, | n_{-}) \,, 
\end{equation}
with $x$ defining the plus-momentum of the quark via $k^{+} = x P^{+}$.
Recently, a lot of work has been devoted to the experimental investigation 
of $k_{T}$-dependent parton distributions --- determined through the 
correlator in (\ref{e:rel}) --- and fragmentation 
functions~\cite{HERMES_99,HERMES_01,CLAS_03,STAR_03,HERMES_04,COMPASS_05}.
Most of these studies have focussed on so-called time-reversal odd (T-odd) 
correlation functions which typically give rise to single spin asymmetries.
Also on the theoretical side there has been a tremendous activity in this 
field of research during the past years comprising conceptual 
(see, e.g., Refs.~\cite{brodsky_02a,collins_02,ji_02,belitsky_02,metz_02,burkardt_02,pobylitsa_03,boer_03a,collins_03,burkardt_04,ji_04a,bomhof_04,collins_04a,ji_05}) 
and phenomenological work (see, e.g., Refs.~\cite{boer_02,efremov_03,bacchetta_03,gamberg_03,boer_03b,dalesio_04,efremov_04a,efremov_04b,anselmino_05,schmidt_05}).
Because many of the mentioned studies are dealing with subleading twist 
(twist-3) effects it is important to have a complete description of the
correlator (\ref{e:rel}) including the twist-3 level.
In the present work we intend to present such a description for the first 
time.
We also would like to emphasize that the totally unintegrated correlator in
Eq.~(\ref{e:corr1}) should not merely be considered as a mathematical object, 
but may in fact be used in the description of hard processes, in which it is 
appropriate to not integrate upon the minus-momentum of the 
quark~\cite{collins_04b}.

Our work is mainly based on the crucial observation made in 
Ref.~\cite{goeke_03} according to which the direction of the Wilson line 
in (\ref{e:corr1}), specified by the light-cone vector $n_{-}$, leads to 
more terms in the decomposition than the ones considered 
in~\cite{ralston_79,mulders_95}.
However, Ref.~\cite{goeke_03} contains only the spin-independent part 
of the correlator (\ref{e:corr1}) explicitly, even though certain 
spin-dependent terms were used in order to derive the violation of 
three specific relations (so-called Lorentz invariance relations) 
between forward twist-3 parton distributions and moments of 
$k_{T}$-dependent parton distributions (see also 
Refs.~\cite{kundu_01,schlegel_04}).
In fact, also the spin-independent part given in~\cite{goeke_03} was not
entirely complete which has subsequently been corrected in 
Ref.~\cite{bacchetta_04}.
It is quite interesting that the one additional structure advocated 
in~\cite{bacchetta_04} implies also a new structure (associated with
a new twist-3 parton distribution, called $g^{\perp}$ in 
Ref.~\cite{bacchetta_04}) on the level of the $k_{T}$-dependent 
correlator in Eq.~(\ref{e:rel}).
In Ref.~\cite{metz_04} the existence of $g^{\perp}$ was already 
anticipated based on a calculation of the single spin asymmetry
$A_{LU}$ (longitudinally polarized lepton beam and unpolarized target) 
for semi-inclusive DIS in the framework of a spectator model
(see also Ref.~\cite{afanasev_03}).

In the present work we want to give the complete structure of the 
correlator in Eq.~(\ref{e:corr1}) for a spin-$\frac{1}{2}$ hadron 
including all terms generated by the presence of the Wilson line.
We find as a particular consequence two new T-odd parton distributions
that appear at twist-3 level in the correlator (\ref{e:rel}).
Altogether the twist-3 part of (\ref{e:rel}) contains 16 parton 
distributions and shows a high degree of symmetry.
\\[0.5cm]
\noindent
{\bf 2.}~We start by specifying the Wilson line that appears in 
Eq.~(\ref{e:corr1}),
\begin{equation} \label{e:wilson1}
 {\cal W}(0, \xi|n_{-}) =
 [0,0,\vec{0}_{T} ; 0,\infty,\vec{0}_{T}]
 \mbox{} \times
 [0,\infty,\vec{0}_{T};\xi^+,\infty,\vec{\xi}_{T}]
 \mbox{} \times
 [\xi^+,\infty,\vec{\xi}_{T};\xi^+,\xi^-,\vec{\xi}_{T}] \,,
\end{equation}
where $[a^+,a^-,\vec{a}_{T};b^+,b^-,\vec{b}_{T}]$ denotes a gauge link 
connecting the points $a^\mu=(a^+,a^-,\vec{a}_{T})$ and 
$b^\mu=(b^+,b^-,\vec{b}_{T})$ along a straight line.
It is important to note that the contour in Eq.~(\ref{e:wilson1}) not 
only depends on the coordinates of the initial and final points but 
also on the light-cone direction $n_{-}$, which is opposite to the 
direction of the target momentum~\cite{goeke_03}.
The path is chosen such that, upon integration over the minus-momentum
of the quark, it leads to a proper definition of the correlator in 
(\ref{e:rel}) as given in 
Refs.~\cite{collins_81,collins_02,ji_02,belitsky_02,boer_03a}.
The choice of the contour depends on the process under 
consideration~\cite{collins_02}. 
Here we restrict ourselves to the case of semi-inclusive DIS, but all 
our arguments hold as well for other processes like Drell-Yan.
It has been pointed out~\cite{collins_81,collins_03} that in general 
light-like Wilson lines as used in (\ref{e:wilson1}) can lead to 
divergences, which can be avoided, however, by adopting a near light-cone 
direction.
Again, our general reasoning remains valid if we use such a direction 
instead of $n_{-}$.

To write down the most general expression of the correlator 
in~(\ref{e:corr1}), we impose the following constraints due to hermiticity 
and parity,
\begin{eqnarray}
\Phi^{\dagger}(P,k,S | n_{-}) & = &
 \gamma_0 \Phi(P,k,S | n_{-}) \gamma_0 \,,
\\
\Phi(P,k,S | n_{-}) & = &
 \gamma_0 \, \Phi(\bar P , \bar k, -\bar S | \bar n_{-}) \gamma_0 \,,
\end{eqnarray}
where $\bar{P}^{\mu} = (P^0 , -\vec{P})$, etc. 
In the case of the correlators (\ref{e:corr1}),(\ref{e:rel}) time-reversal 
does not give an additional constraint~\cite{collins_02}.
To avoid redundant terms in the decomposition we make use of the identity
\begin{equation} \label{e:epsilon}
g^{\alpha\beta} \varepsilon^{\mu\nu\rho\sigma} =
   g^{\mu\beta} \varepsilon^{\alpha\nu\rho\sigma} 
+  g^{\nu\beta} \varepsilon^{\mu\alpha\rho\sigma}
+  g^{\rho\beta} \varepsilon^{\mu\nu\alpha\sigma}
+  g^{\sigma\beta} \varepsilon^{\mu\nu\rho\alpha} \,.
\end{equation}
With these ingredients it is possible to obtain the general form of
the correlator in Eq.~(\ref{e:corr1}).
One ends up with 32 matrix structures multiplied by scalar functions 
($A_i$, $B_i$),
\begin{eqnarray} \label{e:corr1_res}
\Phi (P,k,S|n) &  = &
M A_{1} 
+ \pslash A_{2} 
+ \kslash A_{3}
+ \frac{i}{2M} [\pslash,\kslash] \, A_{4}  
+ i (k \cdot S) \gamma_5 \, A_{5}
+ M \sslash \gamma_5 \, A_{6}
\\
& &
+ \frac{k \cdot S}{M} \pslash \gamma_5 \, A_{7}
+ \frac{k \cdot S}{M} \kslash \gamma_5 \, A_{8}
+ \frac{[\pslash,\sslash]}{2} \gamma_5 \, A_{9}
+ \frac{[\kslash,\sslash]}{2} \gamma_5 \, A_{10}
\nonumber \\
& & 
+ \frac{(k \cdot S)}{2M^2} [\pslash,\kslash] \gamma_5 \, A_{11}
+ \frac{1}{M} \varepsilon^{\mu\nu\rho\sigma} 
  \gamma_{\mu} P_{\nu} k_{\rho} S_{\sigma} \, A_{12} 
\nonumber \\
& & 
+ \frac{M^2}{P \cdot n_{-}} \nslash \, B_{1}
+ \frac{i M}{2 P \cdot n_{-}} [\pslash,\nslash ] \, B_{2}
+ \frac{i M}{2 P \cdot n_{-}} [\kslash,\nslash ] \, B_{3}
\nonumber \\
& &
+ \frac{1}{P \cdot n_{-}} \varepsilon^{\mu\nu\rho\sigma}
  \gamma_{\mu} \gamma_5 P_{\nu} k_{\rho} n_{-\sigma} \, B_{4} 
\nonumber \\
& & 
+ \frac{1}{P \cdot n_{-}} \varepsilon^{\mu\nu\rho\sigma}
  P_{\mu} k_{\nu} n_{-\rho} S_{\sigma} \, B_{5}
+ \frac{i M^2}{P \cdot n_{-}} (n_{-} \cdot S) \gamma_5 \, B_{6} 
\nonumber \\
& &
+ \frac{M}{P \cdot n_{-}} \varepsilon^{\mu\nu\rho\sigma}
  \gamma_{\mu} P_{\nu} n_{-\rho} S_{\sigma} \, B_{7}
+ \frac{M}{P \cdot n_{-}} \varepsilon^{\mu\nu\rho\sigma}
  \gamma_{\mu} k_{\nu} n_{-\rho} S_{\sigma} \, B_{8}
\nonumber \\
& &
+ \frac{(k \cdot S)}{M (P \cdot n_{-})} \varepsilon^{\mu\nu\rho\sigma}
  \gamma_{\mu} P_{\nu} k_{\rho} n_{-\sigma} \, B_{9}
+ \frac{M (n_{-} \cdot S)}{(P \cdot n_{-})^2} \varepsilon^{\mu\nu\rho\sigma}
  \gamma_{\mu} P_{\nu} k_{\rho} n_{-\sigma} \, B_{10}
\nonumber \\
& &
+ \frac{M}{P \cdot n_{-}} (n_{-} \cdot S) \pslash \gamma_5 \, B_{11}
+ \frac{M}{P \cdot n_{-}} (n_{-} \cdot S) \kslash \gamma_5 \, B_{12}
\nonumber \\
& & 
+ \frac{M}{P \cdot n_{-}} (k \cdot S) \nslash \gamma_5 \, B_{13}
+ \frac{M^3}{(P \cdot n_{-})^2} (n_{-} \cdot S) \nslash \gamma_5 \, B_{14}
\nonumber \\
& & 
+ \frac{M^2}{2 P \cdot n_{-}} [\nslash,\sslash] \gamma_5 \, B_{15}
+ \frac{(k \cdot S)}{2 P \cdot n_{-}} [\pslash,\nslash] \gamma_5 \, B_{16}
+ \frac{(k \cdot S)}{2 P \cdot n_{-}} [\kslash,\nslash] \gamma_5 \, B_{17}
\nonumber \\
& & 
+ \frac{(n_{-} \cdot S)}{2 P \cdot n_{-}} [\pslash,\kslash] \gamma_5 \, B_{18}
+ \frac{M^2 (n_{-} \cdot S)}{2 (P \cdot n_{-})^2} [\pslash,\nslash] \gamma_5 \, B_{19}
+ \frac{M^2 (n_{-} \cdot S)}{2 (P \cdot n_{-})^2} [\kslash,\nslash] \gamma_5 \, B_{20} \,.
\nonumber
\end{eqnarray}
The first twelve structures that are multiplied by the amplitudes $A_i$ were 
already written down for the corresponding fragmentation correlator in 
Ref.~\cite{mulders_95}. 
(See Ref.~\cite{boer_97} in the case of parton distributions.)
These terms constitute a complete decomposition as long as the Wilson line is 
neglected. 
They give a sufficient parameterization if the correlator is evaluated in some 
model of non-perturbative QCD which does not contain gluonic degrees of 
freedom.

The spin-independent terms associated with the $n_{-}$-dependence and the
amplitudes $B_{1,2,3}$ were given in~\cite{goeke_03}, while the $B_{4}$-term 
can be found for the first time in~\cite{bacchetta_04}.
The remaining 16 $B$-terms are relevant once the target spin is involved.
Note that in order to specify the Wilson line in Eq.~(\ref{e:wilson1}) a 
rescaled vector $\lambda n_{-}$ with some parameter $\lambda$ could be used
instead of $n_{-}$.
By construction, the terms in (\ref{e:corr1_res}) are not affected by such 
a rescaling.
The various factors of the target mass $M$ are introduced in order to assign
the same mass dimension to all scalar amplitudes. 
Finally, we mention that the following twelve amplitudes are associated with 
T-odd matrix structures:
$A_{4}$, $A_{5}$, $A_{12}$, $B_{2}$, $B_{3}$, $B_{4}$, $B_{5}$, $B_{6}$,
$B_{7}$, $B_{8}$, $B_{9}$, $B_{10}$.
\\[0.5cm]
\noindent
{\bf 3.}~We now focus our attention on the $k_{T}$-dependent correlator 
in Eq.~(\ref{e:rel}),
\begin{equation} \label{e:corr2}
\Phi_{ij}(x,\vec{k}_{T},S)  =
\int \frac{d \xi^- \, d^2 \vec{\xi}_{T}}{(2 \pi)^3} \,
 e^{i (k^+ \xi^- -\vec{k}_{T} \cdot \vec{\xi}_{T})} \,
 \langle P,S \, | \, \bar{\psi}_j(0) \,
 {\cal W}_1(0 , \xi) \,
 \psi_i(\xi) \, | \, P,S \rangle \biggr|_{\xi^+=0} \,,
\end{equation}
which can be derived from the general result (\ref{e:corr1_res}) in a 
straightforward manner.
The Wilson line in this correlator is connected to the one in (\ref{e:wilson1})
through
\begin{equation}
{\cal W}_1(0 , \xi)={\cal W}(0, \xi|n_{-})\biggr|_{\xi^+=0} \,. 
\end{equation}
We will specify the $k_{T}$-dependent correlator in (\ref{e:corr2}) in terms
of all possible Dirac traces given by 
\begin{eqnarray}
\Phi^{[\Gamma]}(x,\vec{k}_{T},S) & \equiv & 
 \frac{1}{2} \, \textrm{Tr} \Big (\Phi(x,\vec{k}_{T},S) \, \Gamma \Big)
\\
& = & \int \frac{d \xi^- \, d^2 \vec{\xi}_{T}}{2 (2 \pi)^3} \,
 e^{i (k^+ \xi^- - \vec{k}_{T} \cdot \vec{\xi}_{T})} \,
 \langle P,S \, | \, \bar{\psi}_j(0) \, \Gamma \,
 {\cal W}_1(0 , \xi) \,
 \psi_i(\xi) \, | \, P,S \rangle \biggr|_{\xi^+=0} \, .
\nonumber
\end{eqnarray}
These traces immediately provide the definition of the various 
$k_{T}$-dependent parton distributions.
In order to have a twist-classification it is convenient to use the
Sudakov decomposition of the four-vectors in (\ref{e:corr1_res}),
\begin{eqnarray}
P^{\mu} & = & P^{+} n_{+}^{\mu} + \frac{M^2}{2 P^{+}} n_{-}^{\mu} \,,
\\
k^{\mu} & = & x P^{+} n_{+}^{\mu} + k^{-} n_{-}^{\mu} + k_{T}^{\mu} \,,
\vphantom{\frac{1}{1}}
\\
S^{\mu} & = & \lambda \frac{P^{+}}{M} n_{+}^{\mu} 
            - \lambda \frac{M}{2 P^{+}} n_{-}^{\mu} + S_{T}^{\mu} \,,
\end{eqnarray} 
with $k_{T}^{\mu} = (0,0,\vec{k}_{T})$ and 
$S_{T}^{\mu} = (0,0,\vec{S}_{T})$.
The two light-like vectors $n_{-}$, $n_{+}$ satisfy the usual conditions
$n_{-}^2 = n_{+}^2 = 0$ and $n_{-} \cdot n_{+} =1$.
We consider $P^{+}$ as the large component of the target momentum.
This input, together with the relation (\ref{e:rel}), is sufficient to obtain 
the final result for the $k_{T}$-dependent correlator.

For the sake of completeness and of later comparison we start with the result 
for the twist-2 case, which has already been given in the 
literature~\cite{mulders_95,boer_97},
\begin{eqnarray} \label{e:1}
\Phi^{[\gamma^+]} & = &
 f_1(x,\vec{k}_{T}^{2}) 
 - \frac{\varepsilon_{T}^{ij} k_{Ti} S_{Tj}}{M} \, 
   f_{1T}^{\perp}(x,\vec{k}_T^{2}) \,,
\\
\Phi^{[\gamma^+ \gamma_5]} & = &
 \lambda \, g_{1L}(x,\vec{k}_T^{2})
 + \frac{\vec{k}_{T} \cdot \vec{S}_{T}}{M} \, 
   g_{1T}(x,\vec{k}_T^{2}) \,,
\\ \label{e:spi5}
\Phi^{[i\sigma^{+i}\gamma_5]} & = &
 S_{T}^{i} \, h_{1T}(x,\vec{k}_T^{2})
 + \frac{k_{T}^{i}}{M} \bigg( \lambda \, h_{1L}^{\perp}(x,\vec{k}_T^{2}) 
   + \frac{\vec{k}_{T} \cdot \vec{S}_{T}}{M} \, h_{1T}^{\perp}(x,\vec{k}_T^{2}) \bigg)
\\
& & - \frac{\varepsilon_{T}^{ij} k_{Tj}}{M} \, h_{1}^{\perp} (x,\vec{k}_T^{2}) \,.
\nonumber
\end{eqnarray}
Here we use the definition $\varepsilon_{T}^{ij} = \varepsilon^{-+ij}$ and the 
standard notation $\sigma^{\mu\nu} = i [\gamma^{\mu},\gamma^{\nu}] / 2$.
All eight twist-2 parton distributions are given by $k^{-}$-integrals of
certain linear combinations of the scalar amplitudes in (\ref{e:corr1_res}).
For brevity we refrain from listing these relations here.
The functions $f_{1T}^{\perp}$ (Sivers function~\cite{sivers_89}) and 
$h_{1}^{\perp}$~\cite{boer_97} are T-odd and have recently attracted an enormous
interest because they are considered to be at the origin of the observed interesting 
single spin phenomena in certain hard processes.
If the correlator is integrated upon $k_T$ only three functions (the forward
unpolarized, helicity and transversity distribution of a quark) survive.

In the twist-3 case, characterized through a suppression by one power in $P^{+}$, 
we find
\begin{eqnarray}
\Phi^{[1]} & = &
 \frac{M}{P^+} \bigg[ e(x,\vec{k}_{T}^{2}) 
 - \frac{\varepsilon_{T}^{ij} k_{Ti} S_{Tj}}{M} \, 
   e_{T}^{\perp}(x,\vec{k}_T^{2}) \bigg] ,
\\
\Phi^{[i \gamma_5]} & = &
 \frac{M}{P^+} \bigg[ \lambda \, e_{L}(x,\vec{k}_T^{2})
 + \frac{\vec{k}_{T} \cdot \vec{S}_{T}}{M} \, 
   e_{T}(x,\vec{k}_T^{2}) \bigg] ,
\\ \label{e:gi}
\Phi^{[\gamma^{i}]} & = &
 \frac{M}{P^+} \bigg[ \frac{k_{T}^{i}}{M} \bigg( f^{\perp}(x,\vec{k}_{T}^{2}) 
 - \frac{\varepsilon_{T}^{jk} k_{Tj} S_{Tk}}{M} \, 
   f_{T}^{\perp\prime}(x,\vec{k}_T^{2}) \bigg)
\\
& & \hspace{0.8cm} 
 + \frac{\varepsilon_{T}^{ij} k_{Tj}}{M} \bigg( 
   \lambda \, f_{L}^{\perp}(x,\vec{k}_T^{2})
 + \frac{\vec{k}_{T} \cdot \vec{S}_{T}}{M} \, 
   f_{T}^{\perp}(x,\vec{k}_T^{2}) \bigg) \bigg] ,
\nonumber \\ \label{e:gi5}
\Phi^{[\gamma^{i}\gamma_5]} & = &
 \frac{M}{P^+} \bigg[ S_{T}^{i} \, g_{T}^{\prime}(x,\vec{k}_T^{2})
 + \frac{k_{T}^{i}}{M} \bigg( \lambda \, g_{L}^{\perp}(x,\vec{k}_T^{2}) 
   + \frac{\vec{k}_{T} \cdot \vec{S}_{T}}{M} \, g_{T}^{\perp}(x,\vec{k}_T^{2}) \bigg)
\\
& & \hspace{0.8cm} 
 - \frac{\varepsilon_{T}^{ij} k_{Tj}}{M} \, g^{\perp} (x,\vec{k}_T^{2}) \bigg] ,
\nonumber \\
\Phi^{[i\sigma^{ij}\gamma_5]} & = &
 \frac{M}{P^+} \bigg[ \frac{S_{T}^{i} k_{T}^{j} - k_{T}^{i} S_{T}^{j}}{M} \,
   h_{T}^{\perp}(x,\vec{k}_T^{2})
 - \varepsilon_{T}^{ij} \, h(x,\vec{k}_T^{2}) \bigg] ,
\\
\Phi^{[i\sigma^{+-}\gamma_5]} & = &
 \frac{M}{P^+} \bigg[ \lambda \, h_{L}(x,\vec{k}_T^{2})
 + \frac{\vec{k}_{T} \cdot \vec{S}_{T}}{M} \, 
   h_{T}(x,\vec{k}_T^{2}) \bigg] .
\end{eqnarray}
The twist-4 result, which is basically a copy of the twist-2 case, reads
\begin{eqnarray}
\Phi^{[\gamma^-]} & = &
 \frac{M^2}{(P^+)^2}\bigg[ f_3(x,\vec{k}_{T}^{2}) 
 - \frac{\varepsilon_{T}^{ij} k_{Ti} S_{Tj}}{M} \, 
   f_{3T}^{\perp}(x,\vec{k}_T^{2}) \bigg] ,
\\
\Phi^{[\gamma^- \gamma_5]} & = &
  \frac{M^2}{(P^+)^2}\bigg[ \lambda \, g_{3L}(x,\vec{k}_T^{2})
 + \frac{\vec{k}_{T} \cdot \vec{S}_{T}}{M} \, 
   g_{3T}(x,\vec{k}_T^{2}) \bigg] ,
\\ \label{e:smi5}
\Phi^{[i\sigma^{-i}\gamma_5]} & = &
  \frac{M^2}{(P^+)^2}\bigg[ S_{T}^{i} \, h_{3T}(x,\vec{k}_T^{2})
 + \frac{k_{T}^{i}}{M} \bigg( \lambda \, h_{3L}^{\perp}(x,\vec{k}_T^{2}) 
   + \frac{\vec{k}_{T} \cdot \vec{S}_{T}}{M} \, h_{3T}^{\perp}(x,\vec{k}_T^{2}) \bigg)
\\
& & \hspace{1.2cm}
 - \frac{\varepsilon_{T}^{ij} k_{Tj}}{M} \, h_{3}^{\perp} (x,\vec{k}_T^{2}) \bigg] .
\nonumber
\end{eqnarray}
The twist-4 case is of course only of academic interest but is included for  
completeness. 
We would like to add several points:
\begin{enumerate}
\item In total there are 32 $k_{T}$-dependent parton distributions which exactly
 agrees with the number of the independent amplitudes in Eq.~(\ref{e:corr1_res}).
 This result seems non-trivial to us for the following reason:
 If the same calculation is performed neglecting the $n_{-}$-dependent terms in
 (\ref{e:corr1_res}) then the number of structures/functions on the level of
 the $k_{T}$-dependent correlator is larger than the number of the amplitudes 
 $A_{i}$.
 This feature gives rise to the Lorentz invariance relations between certain 
 parton distributions~\cite{mulders_95,boer_97}.
 In a gauge theory, however, these relations no longer hold.
\item At twist-3 there appear 16 functions, where 8 of them 
 ($e_{T}^{\perp}$, $e_{L}$, $e_{T}$, $f_{L}^{\perp}$, 
 $f_{T}^{\perp}$, $f_{T}^{\perp\prime}$, $g^{\perp}$, $h$) are T-odd.
\item The structure of the $k_{T}$-dependent fragmentation correlator
 is completely analogous to the case of parton distributions considered here.
 For fragmentation we refer the reader in particular to~\cite{mulders_95}.
\item With the exception of $e_{T}^{\perp}$, $f_{T}^{\perp}$, 
 $f_{T}^{\perp\prime}$, $g^{\perp}$ all other twist-3 functions were already 
 given in Ref.~\cite{mulders_95} (for the fragmentation case). 
 As mentioned above, the function $g^{\perp}$ was introcued 
 in~\cite{bacchetta_04}.
 The remaining three parton distributions are discussed here for the first 
 time.
 Actually $\Phi^{[\gamma^i]}$ in~\cite{mulders_95} contains a term of the type
 $\varepsilon_{T}^{ij} S_{Tj} \, f_{T}(x,\vec{k}_{T}^2)$, which is not 
 present in our result (\ref{e:gi}).
 To get maximal symmetry of the final result we have eliminated such a contribution 
 by means of the identity
 \begin{equation}
 \vec{k}_{T}^2 \, \varepsilon_{T}^{ij} S_{Tj}
 = - k_{T}^{i} \, \varepsilon_{T}^{jk} k_{Tj} S_{Tk}
   + \varepsilon_{T}^{ij} k_{Tj} \, \vec{k}_{T} \cdot \vec{S}_{T} \,, 
 \end{equation} 
 which immediately follows from Eq.~(\ref{e:epsilon}).
 The terms associated with the functions $f_{T}^{\perp}$ and 
 $f_{T}^{\perp\prime}$ are absent in~\cite{mulders_95}, which means that
 $\Phi^{[\gamma^i]}$ in that reference contains only three instead of four
 independent functions.
\item In our work the function $g^{\perp}$ in (\ref{e:gi5}) has the opposite 
 sign as compared to Ref.~\cite{bacchetta_04}.
 We propose this sign reversal because in that case the structure of 
 $\Phi^{[\gamma^i\gamma_5]}$ completely coincides with the twist-2 structure
 $\Phi^{[i\sigma^{+i}\gamma_5]}$ in (\ref{e:spi5}).
\item The parton distributions $e_{T}^{\perp}$, $g^{\perp}$ and the independence 
 of the functions $f_{T}^{\perp}$ and $f_{T}^{\perp\prime}$ only appear if the 
 gauge link is taken into account in the unintegrated correlator 
 in Eq.~(\ref{e:corr1_res}).
 All these functions are T-odd, which is consistent with the fact that they 
 vanish once the gauge link is neglected~\cite{collins_92,collins_02}.
\item If the correlation functions in Eqs.(\ref{e:1})--(\ref{e:smi5}) are 
 integrated upon $k_{T}$ one obtains the light-cone correlators 
 $\Phi^{[\Gamma]}(x)$.
 In these objects all T-odd functions have to vanish due to time-reversal
 invariance of QCD~\cite{collins_92}, which implies the following constraints:
 \begin{eqnarray}
 \int d^2 \vec{k}_{T} \, e_{L}(x,\vec{k}_{T}^2) & = & 0 \,,
 \\
 \int d^2 \vec{k}_{T} \, \vec{k}_{T}^2 
 \Big( f_{T}^{\perp}(x,\vec{k}_{T}^2) 
     + f_{T}^{\perp\prime}(x,\vec{k}_{T}^2) \Big) & = & 0 \,,
 \\ 
\int d^2 \vec{k}_{T} \, h(x,\vec{k}_{T}^2) & = & 0 \,.
 \end{eqnarray}
 Such relations do not hold in the case of the corresponding fragmentation 
 functions.
\item The new functions appear in transverse momentum dependent semi-inclusive
 DIS and in the unintegrated Drell-Yan process at subleading twist.
 To be specific, in semi-inclusive DIS $e_{T}^{\perp}$ enters the double 
 polarized cross section $\sigma_{LT}$ (multiplied with the Collins function), 
 while $f_{T}^{\perp}$ and $f_{T}^{\perp\prime}$ enter $\sigma_{UT}$
 (multiplied with the unpolarized fragmentation function $D_{1}$).
 It is beyond the scope of this article to give a complete (parton model)
 description of these observables up to twist-3, because one has to deal
 also with quark-gluon-quark matrix elements. 
 (In this context see, e.g., Refs.~\cite{mulders_95,boer_03a}.) 
\end{enumerate}
\noindent
{\bf 4.}~In summary, we have derived the general structure of the quark-quark 
correlation function for a spin-$\frac{1}{2}$ hadron.
In order to obtain a full parameterization of the correlator in QCD
it is crucial to consider also the dependence on an additional light-like vector 
specifying the direction of the Wilson line, which ensures color gauge invariance 
of the correlator.
We have used the result to write down the most general form of the 
$k_{T}$-dependent quark-quark correlator $\Phi(x,\vec{k}_{T},S)$ 
appearing in the description of various hard scattering processes.
Our final result for this correlator shows a high degree of symmetry.
In particular, we have found two new $k_{T}$-dependent T-odd parton distributions 
at subleading twist. 
\\[0.6cm]
\noindent
{\bf Acknowledgements:}
We are grateful to P.V. Pobylitsa and M.V. Polyakov for discussions.
The work of M.S. has been supported by the Graduiertenkolleg
``Physik der Elementarteilchen an Beschleunigern und im Universum.''
The work has also been partially supported by the Verbundforschung (BMBF)
and the Transregio/SFB Bochum-Bonn-Giessen.
This research is part of the EU Integrated Infrastructure Initiative
Hadronphysics Project under contract number RII3-CT-2004-506078.



\begin{thebibliography}{99}
\bibitem{ralston_79}
J.P.~Ralston and D.E.~Soper,
Nucl. Phys. {\bf B152}, 109 (1979).

\bibitem{mulders_95}
P.J.~Mulders and R.D.~Tangerman,
Nucl. Phys. {\bf B461}, 197 (1996), [Erratum-ibid. {\bf B484} (1997) 538].

\bibitem{goeke_03}
K.~Goeke, A.~Metz, P.V.~Pobylitsa, and M.V.~Polyakov,
Phys. Lett. B {\bf 567} 27 (2003).

\bibitem{bacchetta_04}
A.~Bacchetta, P.J.~Mulders, and F.~Pijlman,
Phys. Lett. B {\bf 595}, 309 (2004).

\bibitem{HERMES_99}
A.~Airapetian {\it et al.} [HERMES Collaboration],
Phys. Rev. Lett. {\bf 84}, 4047 (2000).

\bibitem{HERMES_01}
A.~Airapetian {\it et al.} [HERMES Collaboration],
Phys. Rev. D {\bf 64}, 097101 (2001).

\bibitem{CLAS_03}
H.~Avakian {\it et al.} [CLAS Collaboration],
Phys. Rev. D {\bf 69}, 112004 (2004).

\bibitem{STAR_03}
J.~Adams {\it et al.} [STAR Collaboration],
Phys. Rev. Lett. {\bf 92}, 171801 (2004).

\bibitem{HERMES_04}
A.~Airapetian {\it et al.} [HERMES Collaboration],
Phys. Rev. Lett. {\bf 94}, 012002 (2005).

\bibitem{COMPASS_05}
V.Y.~Alexakhin {\it et al.} [COMPASS Collaboration],
hep-ex/0503002.

\bibitem{brodsky_02a}
S.J.~Brodsky, D.S.~Hwang, and I.~Schmidt,
Phys. Lett. B {\bf 530}, 99 (2002).

\bibitem{collins_02}
J.C.~Collins,
Phys. Lett. B {\bf 536}, 43 (2002).

\bibitem{ji_02}
X.~Ji and F.~Yuan,
Phys. Lett. B {\bf 543}, 66 (2002).

\bibitem{belitsky_02}
A.V.~Belitsky, X.~Ji, and F.~Yuan,
Nucl. Phys. {\bf B656}, 165 (2003).

\bibitem{metz_02}
A.~Metz,
Phys. Lett. B {\bf 549}, 139 (2002).

\bibitem{burkardt_02}
M.~Burkardt,
Phys. Rev. D {\bf 66}, 114005 (2002).

\bibitem{pobylitsa_03}
P.V.~Pobylitsa,
arXiv:hep-ph/0301236.

\bibitem{boer_03a}
D.~Boer, P.J.~Mulders, and F.~Pijlman,
Nucl. Phys. {\bf B667}, 201 (2003).

\bibitem{collins_03}
J.C.~Collins,
Acta Phys. Polon. {\bf B34}, 3103 (2003).

\bibitem{burkardt_04}
M.~Burkardt,
Phys. Rev. D {\bf 69}, 091501 (2004).

\bibitem{ji_04a}
X.~Ji, J.P.~Ma, and F.~Yuan,
Phys. Rev. D {\bf 71}, 034005 (2005).

\bibitem{bomhof_04}
C.J.~Bomhof, P.J.~Mulders, and F.~Pijlman,
Phys. Lett. B {\bf 596}, 277 (2004).

\bibitem{collins_04a}
J.C.~Collins and A.~Metz,
Phys. Rev. Lett. {\bf 93}, 252001 (2004).

\bibitem{ji_05}
X.~Ji, J.P.~Ma, and F.~Yuan,
arXiv:hep-ph/0503015.

\bibitem{boer_02}
D.~Boer, S.J.~Brodsky, and D.S.~Hwang,
Phys. Rev. D {\bf 67}, 054003 (2003).

\bibitem{efremov_03}
A.V.~Efremov, K.~Goeke, and P.~Schweitzer,
Eur. Phys. J. {\bf C32}, 337 (2003).

\bibitem{bacchetta_03}
A.~Bacchetta, A.~Schaefer, and J.J.~Yang,
Phys. Lett. B {\bf 578}, 109 (2004).

\bibitem{gamberg_03}
L.P.~Gamberg, D.S.~Hwang, and K.~A.~Oganessyan,
Phys. Lett. B {\bf 584}, 276 (2004).

\bibitem{boer_03b}
D.~Boer and W.~Vogelsang,
Phys. Rev. D {\bf 69}, 094025 (2004).

\bibitem{dalesio_04}
U.~D'Alesio and F.~Murgia,
Phys. Rev. D {\bf 70}, 074009 (2004). 

\bibitem{efremov_04a}
A.V.~Efremov, K.~Goeke, S.~Menzel, A.~Metz, and P.~Schweitzer,
Phys. Lett. B {\bf 612}, 233 (2005).

\bibitem{efremov_04b}
A.V.~Efremov, K.~Goeke, and P.~Schweitzer,
hep-ph/0412420.

\bibitem{anselmino_05}
M.~Anselmino, M.~Boglione, U.~D'Alesio, A.~Kotzinian, F.~Murgia,
and A.~Prokudin,
hep-ph/0501196.

\bibitem{schmidt_05}
I.~Schmidt, J.~Soffer, and J.J.~Yang,
arXiv:hep-ph/0503127.

\bibitem{collins_04b}
J.C.~Collins and X.~Zu,
hep-ph/0411332.

\bibitem{kundu_01}
R.~Kundu and A.~Metz,
Phys. Rev. D {\bf 65}, 014009 (2002).

\bibitem{schlegel_04}
M.~Schlegel and A.~Metz,
hep-ph/0406289.

\bibitem{metz_04}
A.~Metz and M.~Schlegel,
Eur. Phys. J. {\bf A22}, 489 (2004).

\bibitem{afanasev_03}
A.~Afanasev and C.E.~Carlson,
hep-ph/0308163.

\bibitem{collins_81}
J.C.~Collins and D.E.~Soper,
Nucl. Phys. {\bf B194}, 445 (1982).

\bibitem{boer_97}
D.~Boer and P.J.~Mulders,
Phys. Rev. D {\bf 57}, 5780 (1998).

\bibitem{sivers_89}
D.W.~Sivers,
Phys. Rev. D {\bf 41}, 83 (1990); Phys. Rev. D {\bf 43}, 261 (1991). 

\bibitem{collins_92}
J.C.~Collins,
Nucl. Phys. {\bf B396}, 161 (1993).

\end{thebibliography}
\end{document}